# Threshold Asymmetric Conditional Autoregressive Range (TACARR) Model

**Isuru Ratnayake[1*], V.A. Samaranayake [2]**

1. Department of Biostatistics & Data Science, University of Kansas Medical Center, USA
2. Department of Mathematics & Statistics, Missouri University of Science & Technology, USA

## Abstract

This paper introduces a Threshold Asymmetric Conditional Autoregressive Range (TACARR) formulation for modeling the daily price ranges of financial assets. It is assumed that the process generating the conditional expected ranges at each time point switches between two regimes, labeled as upward market and downward market states. The disturbance term of the error process is also allowed to switch between two distributions depending on the regime. It is assumed that a self-adjusting threshold component that is driven by the past values of the time series determines the current market regime. The proposed model is able to capture aspects such as asymmetric and heteroscedastic behavior of volatility in financial markets. The proposed model is an attempt at addressing several potential deficits found in existing price range models such as the Conditional Autoregressive Range (CARR), Asymmetric CARR (ACARR), Feedback ACARR (FACARR) and Threshold Autoregressive Range (TARR) models. Parameters of the model are estimated using the Maximum Likelihood (ML) method. A simulation study shows that the ML method performs well in estimating the TACARR model parameters. The empirical performance of the TACARR model was investigated using IBM index data and results show that the proposed model is a good alternative for in-sample prediction and out-of-sample forecasting of volatility.

*Corresponding author E-mail address: rratnayake@ku.edu

# 1. Introduction

Modelling economic volatility is indispensable for obtaining a better understanding of the dynamics of financial markets. Financial volatility of asset prices has been studied extensively in the financial and econometric literature over past few decades. Models that allow for asymmetric behavior with respect to positive and negative market shocks have been of significant interest. Recent interest has also focused on measuring volatility that is present during a trading day, say using the daily range, rather than manifested through closing price changes. Herein we propose a range-based volatility model that contains the above features and switches regimes dynamically, based on past values of the time series. Prior to introducing the proposed model, a discussion of the relevant literature is presented in the following.

Engle (1982) proposed the Autoregressive Conditional Heteroscedasticity (ARCH) model to address the complexities of time-varying volatility and volatility clustering in financial time series. Bollerslev (1986) proposed the Generalized Autoregressive Conditional Heteroscedasticity (GARCH) formulation, which remains one of the most popular volatility models up to date. The GARCH formulation models the conditional volatility as a function of lagged squared returns and past conditional variances. Since both aforementioned formulations focus on modeling price returns, they can be identified as examples of return-based volatility models.

In many financial time series applications, standard deviation is the most commonly used measure of stock return volatility. Since the origination of the concept of volatility, researchers have sought alternative ways of measuring it. Parkinson (1980) argued that volatility measures could be calculated using the daily high, daily low, and opening prices of a stock in addition to the traditional closing prices. Parkinson concluded that the range-based method is far superior to the standard methods based on returns. Beckers (1983) tested the validity of different volatility estimators and showed that using the range of a stock price was better at capturing volatility than using the close-to-close changes in asset prices. Kunitomo (1992) added to Parkinson's original result and proposed a new range-based estimator which the author claimed is "ten times more efficient" than the standard volatility estimator. In another study, Alizadeh, Brandt, and Diebold (2002) established that the range-based volatility estimators are highly efficient compared to the classical volatility proxies based on log absolute returns or squared returns.

Several scholars have focused on the alternative approach to modeling volatility and developed theoretical frameworks for range-based models, along with extensive empirical examples of their application. Some examples are the works of Chou (2005, 2006), Brandt and Jones (2006), and Chou and Liu (2010). Chou (2005) introduced the Conditional Autoregressive Range (CARR) model as a special case of Autoregressive Conditional Duration (ACD) model of Engle (1998). The CARR formulation is employed to model price volatility of an asset by considering the range of the log prices for a given fixed time interval such as a trading day. The formulation of the CARR model is similar to that of the standard GARCH model. The main distinction between the two models is that the GARCH uses the rate of return while CARR uses the range as its volatility measure. Chou showed empirically, via out-of-sample forecasting of S&P 500 data, that CARR is an efficient tool for modeling the volatility clustering property compared to the GARCH model. Brandt and Jones (2006) integrated the properties of exponential GARCH (Nelson, 1991) with daily log range data and proposed a range-based Exponential GARCH model. This model has a simple framework, but it is an effective tool for capturing the important characteristics that are present in stock return data such as clustering, volatility asymmetry, and log normality.

Extensive modifications to the CARR model include works of Chiang, Chou, and Wang (2016), who suggested the application of the Lognormal Logarithmic CARR (Lognormal Log CARR) model in the outlier detection process. One major advantage of using a Logarithmic CARR (Log CARR) model is that it allows the relaxing of positivity restrictions on the parameters when calculating conditional expectation. Xie and Wu (2017) modeled the disturbance term in the CARR model using the gamma distribution (GCARR) and showed that the GCARR outperformed Weibull CARR (WCARR) model in its forecasting ability through an empirical study.

Asymmetric volatility refers to the common phenomenon where conditional volatilities show higher fluctuations during downward trends than during upward trends. Yet the CARR model proposed by Chou (2005) used range as the measure of price volatility and treated maximum price and minimum price symmetrically. That is, a high magnitude range due to a large increase in price during a trading day was treated the same as a similar range manifested due to a large decrease in price. However, in the same study, he suggested the CARRX models (namely CARRX-a, and CARRX-b) which include exogenous variables such as (a) lagged return and (b) lagged absolute returns in the conditional mean equation. The purpose of this incorporation was to model one form of asymmetry, namely the leverage effect of Black and Nelson (1991). Chou (2006) presented the Asymmetric CARR (ACARR) model in which both upward and downward price ranges were treated separately, using independent formulations to model each of them separately, with no cross-feedback between them. The upward range was defined as the difference between the maximum price and the opening price, whereas the downward range was defined as the difference between the opening price and the minimum price, all observed within a trading day. Note that these definitions can be extended to periods beyond a day, such as a week, in a similar manner. The ACARR model was extended to the ACARRX model by including exogenous variables such as trading volume (Lamourex and Lastrapes, 1990), lag returns (Black, 1976; Nelson, 1990), or a seasonal factor to count the leverage effect. The Feedback ACARR (FACARR) model proposed by Xie (2018) is a natural extension of the ACARR model. Xie allowed the conditional mean of upward (downward) range to be modeled by incorporating lagged downward (upward) ranges into each sub-model. They showed via extensive empirical studies that the proposed FACARR performed significantly better than the ACARR in both in-sample and out-of-sample forecasting.

All the above models capture the asymmetry in price range data either by introducing a leverage variable or by treating upward and downward price range series separately or allowing limited cross-feedback. An alternative approach to modeling the asymmetric behavior in volatility is to use a threshold model. The idea of threshold models in time series analysis was first introduced by Tong (1978). Tong pointed out the limitations of the linear Gaussian time series models and emphasized the advantage of using nonlinear time series models. He also proposed a nonlinear Threshold Autoregressive (TAR) model. Tsay (1989) introduced a simple but effective method for testing and modeling procedures for TAR models. Threshold ARCH (TARCH) proposed by Zakoian (1991) models the standard deviation conditional on the sign of the previous time periods' returns. Zakoian (1994) improved the existing TARCH model by incorporating the lagged conditional standard deviation and named it as the Threshold GARCH (TGARCH). The GJR-GARCH model developed by Glosten, Jagannathan, and Runkle (1993) and the TGARCH model share notable conceptual similarities. Li and Lam (1995) modeled the asymmetric behavior in the stock returns using a threshold-type ARCH model. Using the Hang Seng index, they showed that the conditional mean of the return series fluctuates according to the ups and downs of the financial market on the previous day. Zhang, Russel, and Tsay (2001) proposed a nonlinear durational model and named it as Threshold Autoregressive Conditional Duration (TACD) to analyze the transaction data. The TACD permits the expected duration to behave nonlinearly based on past durations. Men, Kolkiewicz and Wirjanto (2019) proposed the Threshold Stochastic Conditional Duration (TSCD) model which is an extension to the SCD models originally proposed by Bauwens, Luc and David (2004). In the TSCD model, the authors used either a Gamma or a Weibull distribution to model the innovations. The log conditional duration in the TSCD follows a TAR (1) process, and it switches between two regimes. Chen, Gerlach and Lin (2008) proposed the threshold heteroscedastic models in a range-based setting to analyze the intraday price range. In their paper a nonlinear volatility model for range, named as the Threshold Conditional Autoregressive Range (TARR) model, was introduced. This model switches regimes based on whether or not the previous period exceeds a fixed threshold. They also introduced the TARRX model in which an exogenous variable is used, which is compared against a preset threshold value to determine the regime switching behavior. All the above threshold models switch regimes based on a fixed, predetermined threshold, which has to be selected a priori. Therefore, there is need to develop models where regime switching occurs dynamically, with past data alone determining the switch without a predetermined threshold. The main goal of this paper is to fill this need.

In the following, a parsimonious nonlinear time series model that can capture the heteroscedastic and asymmetric behaviors existing in the financial markets is proposed. We name this process the Threshold Asymmetric Conditional Autoregressive Range (TACARR) model. The TACARR allows the model for the conditional mean to switch between one of the two market regimes (upward market and downward market). In addition, the disturbance term of the model is also assumed to be determined by the current regime the time series is in. In this formulation, the threshold values are self-adjusting, and every time new information arrives it dynamically determines the status of the market regime. Moreover, the proposed TACARR formulation models the heteroscedastic volatility structure and capture the asymmetric behavior with fewer parameters compared to competing models such as the FACARR.

This paper is organized as follows. Section 2 reviews the conditional heteroscedastic range-based CARR, ACARR and TARR models which will be relevant to understanding how the proposed model differs from these. Section 3 introduces the proposed TACARR model and its statistical properties. Following that, in Section 3, we develop maximum likelihood methods to estimate the model parameters for two versions of the proposed model, namely the Exponential TACARR (ETACARR) and Lognormal TACARR (LNTACARR). Section 4 presents the out-of-sample forecasting method and performance evaluation techniques that were employed. Section 5 discusses the results of a simulation study for both ETACARR and LNTACARR models. Section 6 presents the results of an empirical study fitting the proposed TACARR model to IBM stock return data and compares the results with those of several other range-based models. Finally, Section 7 presents concluding remarks.

## 2. Review of CARR, ACARR, FACARR and TARR Models

### 2.1. The Conditional Autoregressive Range Model

Chou (2005) proposed the CARR, which fit the price volatility of an asset by considering range as a measure of price volatility. Let $P_s$ be the logarithmic price of an asset at time point $s \in (t-1, t]$, and let the highest and lowest logarithmic prices of an asset during the interval $(t-1, t]$ be denoted by $P_t^{high}$ and $P_t^{low}$ respectively. Let $R_t$ be the price range defined over the fixed time period $(t-1, t]$ defined by

$$R_t = P_t^{high} - P_t^{low}.$$

Note that it is common for the time interval $(t-1, t]$ to denote the trading period on day *t*, but it can be defined for any period, such as a trading week.

The CARR model of order (*p*, *q*) denoted as CARR (*p*, *q*) is defined as follows:

$$R_t = \lambda_t \varepsilon_t,$$

$$E(R_t | \mathbb{F}_{t-1}) = \lambda_t = \omega + \sum_{i=1}^{p} \alpha_i R_{t-i} + \sum_{j=1}^{q} \beta_j \lambda_{t-j},$$

$$\varepsilon_t \sim i.i.d. \, f(.), E(\varepsilon_t) = 1,$$

$$t = max(p, q), \ldots, T, \text{and}$$

$$0 < \sum_{i=1}^{p} \alpha_i + \sum_{j=1}^{q} \beta_j < 1, \alpha_i > 0, \beta_j > 0.$$

Here, $\lambda_t$ is the expectation of the price range conditional on $\mathbb{F}_{t-1} = \sigma\left[\{R_s\}_{s=1}^{t-1}\right]$, the sigma filed generated by the past ranges up to time *t*-1. The non-negative disturbance term, also known as the standardized range, is denoted by $\varepsilon_t$ and the $\{\varepsilon_t\}$ are assumed to be independent and identically distributed with probability density function $f(.)$ with a non-negative support and unit mean.

## 2.2. The Asymmetric Conditional Autoregressive Range (ACARR) Model

The ACARR model presented by Chou (2006), decompose the range $(R_t)$ series into two components, namely upward range $(R_t^u)$ and downward range $(R_t^d)$. Upward and downward ranges, for the trading period (*t*-1, *t*] on day *t* are expressed in terms of the daily high $(P_t^{high})$, daily low $(P_t^{low})$, and the opening $(P_t^{open})$ logarithmic prices of an asset respectively, as follows:

$$R_t^u = P_t^{high} - P_t^{open},$$
$$R_t^d = P_t^{open} - P_t^{low},$$
$$R_t = R_t^u + R_t^d = P_t^{high} - P_t^{open} + P_t^{open} - P_t^{low} = P_t^{high} - P_t^{low}.$$
(2.1)

Here, the upward range measures the maximum gain or the positive shock to the asset while downward range calculates the minimum gain or the negative impact to the asset price during the trading day *t*.

The CARR model is symmetric because it treats the high and low price symmetrically. However, it is possible to assume that the upward and downward movements exhibit different in their dynamics of the volatility shocks. To allow the asymmetric behavior in price range data, Chou (2006) proposed the ACARR model. ACARR model of order (*p*, *q*) is presented as follows. Let

$$R_t = R_t^u + R_t^d,$$
$$R_t^u = \lambda_t^u \varepsilon_t^u,$$
$$R_t^d = \lambda_t^d \varepsilon_t^d,$$
$$E(R_t^u | \mathbb{F}_{t-1}^u) = \lambda_t^u = \omega^u + \sum_{i=1}^p \alpha_i^u R_{t-i}^u + \sum_{j=1}^q \beta_j^u \lambda_{t-j}^u,$$
$$E(R_t^d | \mathbb{F}_{t-1}^d) = \lambda_t^d = \omega^d + \sum_{i=1}^p \alpha_i^d R_{t-i}^d + \sum_{j=1}^q \beta_j^d \lambda_{t-j}^d,$$
$$\varepsilon_t^u \sim i.i.d. f^u(.), E(\varepsilon_t^u) = 1,$$
$$\varepsilon_t^d \sim i.i.d. f^d(.), E(\varepsilon_t^d) = 1,$$
$$t = max(p,q), \ldots, T,$$
$$0 < \sum_{i=1}^p \alpha_i^u + \sum_{j=1}^q \beta_j^u < 1, \alpha_i^u > 0, \beta_j^u > 0, \text{ and}$$
$$0 < \sum_{i=1}^p \alpha_i^d + \sum_{j=1}^q \beta_j^d < 1, \alpha_i^d > 0, \beta_j^d > 0.$$
(2.2)

Note that $\lambda_t^u \left(= E(R_t^u | \mathbb{F}_{t-1}^u)\right)$ is the mean of the upward range conditional of the $\sigma$-field $\mathbb{F}_{t-1}^u$ generated by the set of random variables $\{R_k^u : k \leq t-1\}$, and $\lambda_t^d \left(= E(R_t^d | \mathbb{F}_{t-1}^d)\right)$ is the mean of the downward range conditional on the $\sigma$-field $\mathbb{F}_{t-1}^d$ generated by $\{R_k : k \leq t-1\}$. The series of disturbance terms of the upward (downward) range model, given by $\{\varepsilon_t^u\}$ $(\{\varepsilon_t^d\})$ are independently and identically distributed as the density function $f^u(.)$ $(f^d(.))$ with unit mean. In addition, the two series of disturbance terms are assumed to be independent of each other. Moreover, the pairs of parameters, $(\omega^u, \omega^d), (\alpha_i^u, \alpha_i^d), (\beta_j^u, \beta_j^d)$ reflect the asymmetric behavior between the upward range and downward ranges.

## 2.3. The Feedback Asymmetric Conditional Autoregressive Range (FACARR) Model

The ACARR model assumes that there is independence between the upward and downward shocks, and Xie (2018) argued against this assumption and presented the FACARR model. This model includes cross-interdependence terms on top of the ACARR setting. Following the same definitions and notations used in Section 2.2, the FACARR model is defined as follows:

$$R_t = R_t^u + R_t^d,$$
$$R_t^u = \lambda_t^u \varepsilon_t^u,$$
$$R_t^d = \lambda_t^d \varepsilon_t^d,$$
$$\lambda_t^u = \omega^u + \sum_{i=1}^{p} \alpha_i^u R_{t-i}^u + \sum_{j=1}^{q} \beta_j^u \lambda_{t-j}^u + \sum_{k=1}^{l} \gamma_k^u R_{t-k}^d,$$
$$\lambda_t^d = \omega^d + \sum_{i=1}^{p} \alpha_i^d R_{t-i}^d + \sum_{j=1}^{q} \beta_j^d \lambda_{t-j}^d + \sum_{k=1}^{l} \gamma_k^d R_{t-k}^u,$$
$$\varepsilon_t^u \sim i.i.d. f^u(.), \ E(\varepsilon_t^u) = 1,$$
$$\varepsilon_t^d \sim i.i.d. f^d(.), \ E(\varepsilon_t^d) = 1,$$
$$t = max(p,q,l), \ldots, T,$$
$$0 < \sum_{i=1}^{p} \alpha_i^u + \sum_{j=1}^{q} \beta_j^u < 1, \alpha_i^u > 0, \beta_j^u > 0, \text{ and}$$
$$0 < \sum_{i=1}^{p} \alpha_i^d + \sum_{j=1}^{q} \beta_j^d < 1, \alpha_i^d > 0, \beta_j^d > 0.$$

(2.3)

In addition to the previous parameter set discussed under the ACARR model in (2.2), FACARR has a new pair of parameters, namely $(\gamma^u, \gamma^d)$, which measures the magnitude and the direction of the lagged upward (downward) range on conditional mean range.

## 2.4. The Threshold Autoregressive Range (TARR) Model

Chen et al. (2008) proposed the Threshold Autoregressive Range (TARR) model that is a range-based threshold heteroskedastic model to analyze price range data. Let $P_s$ be the logarithmic price of an asset at time $s \in (t-1, t]$, and $R_t = P_t^{high} - P_t^{low}$ be the price range for a fixed time interval $(t-1, t]$, then a TARR model with order $(p, q)$ having two regimes is defined as given in the equation (2.4). Generally, the market regimes are defined based on either the information about local market movement such as past values of the price range series or an exogenous variable such as international market behavior, interest rates, and financial index. If the regimes were picked based on an exogenous variable, then the model is called as TARRX. The TARR formulation is as follows:

$$R_t = \lambda_t \varepsilon_t,$$

$$E(R_t \mid \mathbb{F}_{t-1}) = \lambda_t = \begin{cases} \lambda_t^{(r1)} = \omega^{(r1)} + \sum_{i=1}^{p} \alpha_i^{(r1)} R_{t-i} + \sum_{j=1}^{q} \beta_j^{(r1)} \lambda_{t-j} : r_0 \leq Z_{t-d} < r_1 \\ \lambda_t^{(r2)} = \omega^{(r2)} + \sum_{i=1}^{p} \alpha_i^{(r2)} R_{t-i} + \sum_{j=1}^{q} \beta_j^{(r2)} \lambda_{t-j} : r_1 \leq Z_{t-d} < \infty \end{cases},$$

$$\varepsilon_t \sim i.i.d. \, f(.), \, E(\varepsilon_t) = 1,$$

$$t = max(p, q), \ldots, T,$$

$$\omega^{(r1)} > 0, \alpha_i^{(r1)} \geq 0, \beta_j^{(r1)} \geq 0, \text{ and } \sum_{i=1}^{p} \alpha_i^{(r1)} + \sum_{j=1}^{q} \beta_j^{(r1)} < 1,$$

$$\omega^{(r2)} > 0, \alpha_i^{(r2)} \geq 0, \beta_j^{(r2)} \geq 0, \text{ and } \sum_{i=1}^{p} \alpha_i^{(r2)} + \sum_{j=1}^{q} \beta_j^{(r2)} < 1.$$

Here, *r1* represents the market regime 1, while *r2* denotes the market regime 2. The threshold values $r_k$ satisfy $0 \leq r_0 < r_1 < \infty$. The thresholding variable $Z_{t-d}$ with the delay lag $d$ (where $d > 0$) determines the market regime existing at the time *t*. In the two-regime case, if $Z_{t-d} \in [r_0, r_1)$ then market belongs to regime *r1* while if $Z_{t-d} \in [r_1, \infty)$ then it belongs to market regime *r2*. As mentioned above, $Z_{t-d}$ is either a statistic based on previous price information such as $R_{t-d}$ for some lag *d*, or an exogenous variable. Chen et al. (2008) employed the assumption that unconditional volatility is higher in regime 1 (*r1*) than that of the regime 2 (*r2*) to set the threshold value in his two-regime empirical study. Following this reasoning, we employed the constant threshold $\overline{R}$ to determine market regimes when applying TARR model. This threshold value is estimated using the unconditional mean range computed using sample data. For our comparative study, the lagged market price range $R_{t-d}$ is used as the thresholding variable $Z_{t-d}$. The regimes were determined as follows: If $R_{t-d}$ is above the average stock price (i.e., $Z_{t-d} = R_{t-d} \in \left[\overline{R}, \infty\right)$), then market was assumed to belong to the *r1* regime and if the thresholding variable falls below the average stock price (i.e., $Z_{t-d} = R_{t-d} \in \left[0, \overline{R}\right)$), then the market was assumed to belong to the *r2* regime.

## 3. Threshold Asymmetric Conditional Autoregressive Range (TACARR) Model

Here we introduced our proposed TACARR model. Let $\{R_t\}$ be a sequence of price range values for an asset defined over $T$ time intervals such that $t = 1, 2, 3, ..., T$, with $R_t$ calculated as described in Section 2.2. In addition, the upward price range and downward price range components as defined in that section will also be utilized. The proposed model will switch between two regimes based on the past values taken by these components. In contrast to TARR, which requires the pre-selection of an appropriate threshold, the proposed model switches regimes solely based on a comparison of past upward and downward price ranges over a period.

The proposed TACARR model of order (*l*, *p*, *q*) is presented as follows. Let

$$R_t = \lambda_t \varepsilon_t,$$

$$E(R_t | \mathbb{F}_{t-1}) = \lambda_t = \begin{cases} \lambda_t^{(U)} = \omega^{(U)} + \sum_{i=1}^{p} \alpha_i^{(U)} R_{t-i} + \sum_{j=1}^{q} \beta_j^{(U)} \lambda_{t-j} : C_{l,t}^{(U)} \geq C_{l,t}^{(D)} \\ \lambda_t^{(D)} = \omega^{(D)} + \sum_{i=1}^{p} \alpha_i^{(D)} R_{t-i} + \sum_{j=1}^{q} \beta_j^{(D)} \lambda_{t-j} : C_{l,t}^{(U)} < C_{l,t}^{(D)} \end{cases},$$

$$t = max(p, q, l), \ldots, T,$$

$$\varepsilon_t = \begin{cases} \varepsilon_t^{(U)} \sim i.i.d. \, f^{(U)}(.), E(\varepsilon_t^{(U)}) = 1 : C_{l,t}^{(U)} \geq C_{l,t}^{(D)} \\ \varepsilon_t^{(D)} \sim i.i.d. \, f^{(D)}(.), E(\varepsilon_t^{(D)}) = 1 : C_{l,t}^{(U)} < C_{l,t}^{(D)} \end{cases},$$

$$C_{l,t}^{(U)} = \sum_{i=1}^{l} I_i \left[ R_{t-i}^u \geq R_{t-i}^d \right],$$

$$C_{l,t}^{(D)} = \sum_{i=1}^{l} I_i \left[ R_{t-i}^u < R_{t-i}^d \right],$$

$$cov(\varepsilon_t^{(U)}, \varepsilon_t^{(D)}) = 0,$$

$$\omega^{(U)} > 0, \alpha_i^{(U)} \geq 0, \beta_j^{(U)} \geq 0, \text{ and}$$

$$\omega^{(D)} > 0, \alpha_i^{(D)} \geq 0, \beta_j^{(D)} \geq 0. \quad (3.1)$$

Note that $\mathbb{F}_{t-1}$ is the sigma field generated by the information set including range, upward range, and downward range up to time $t$-1. That is, $\mathbb{F}_{t-1} = \sigma\left[\{R_s\}_{s=1}^{t-1}, \{R_s^u\}_{s=1}^{t-1}, \{R_s^d\}_{s=1}^{t-1}\right] = \sigma\left[\{R_s^u\}_{s=1}^{t-1}, \{R_s^d\}_{s=1}^{t-1}\right]$. Also note that in the above formulation, the relationship between $\lambda_t$, the conditional expectation of $R_t$ given $\mathbb{F}_{t-1}$, and past $R_{t-i}$'s and $\lambda_{t-i}$'s depends on the current regime, which in turn is determined by upward and downward range values over past $l$ time periods. Letting the realized market regime $m$ take two possible values $m=(U=$upward market, $D=$downward market), we assume that the innovation terms for market regime $m$, denoted by $\{\varepsilon_t^{(m)}\}$, forms an independent and identically distributed sequence with non-negative support $[0, \infty)$ with density function $f^{(m)}(.)$ such $E\left[\varepsilon_t^{(m)}\right] = 1$ and $cov(\varepsilon_t^{(U)}, \varepsilon_t^{(D)}) = 0$. Note that the $I_i[A_i]$ in equation (3.1) are indicator functions taking the value one if $A_i$ is true and zero otherwise. Observe that equation (3.1) explicitly states that the market condition at time $t$, say $M_t$, is given by

$$M_t = \begin{cases} U : C_{l,t}^{(U)} \geq C_{l,t}^{(D)} \\ D : C_{l,t}^{(U)} < C_{l,t}^{(D)} \end{cases}.$$

(3.2)

In lay terms, the regime is labeled as an upward market regime when in a majority of periods $(t-i-1, t-i]$, $i=1, 2, ..., l$, the upward range exceeded the downward range. The downward market regime is defined in a similar manner. For example, if the time series consists of daily data and $l=5$, then the current day is considered to be in the upward market regime if in three or more of the past five trading days the upward range exceeded the downward range. Otherwise, the current day is designated as being in the downward market. If $l=1$, then the market regime is determined by which of the upward or downward ranges was greater in the previous trading day.

We propose the TACARR model as an alternative to the regular CARR model which assumes a symmetric behavior to upward and downward market conditions. The proposed model can overcome this drawback by permitting the conditional mean to depend nonlinearly on past information contained in the price series itself. The TACARR model is similar to the TARR formulation, with some differences. In TARR, the regime is switched based on the magnitude of the past range, that is the sum of the upward and downward ranges a-rather than a finer examination of the relationship between the two components. Thus, the asymmetry implied by TARR is related to the magnitude of the price swing rather than whether the market was upstreaming or down treading. Of course, in the TARRX model, and exogenous variable that indicates such trends can be incorporated, but our proposed model achieves this using the observed time series data themselves. Moreover, in the TARR formulation, a pre-determined threshold is used, with the regime determined by whether the past range exceeds this threshold or not. In TARRX, the regime is determined by an exogenous variable and a predetermined threshold. In TACARR, such a predetermined and static threshold value is not required. What needs pre-selection in TACARR is the number of past periods, $l$, one uses to determine the regime. The researcher can select one value for $l$ out of several based-on AIC/BIC criteria or on out of sample predictive performance. One drawback of TACARR relative to TARR, however, is that it requires the upward and downward range values, but this is not a serious issue because many price series of assets report starting, maximum, and minimum prices for each trading day. Another difference between TARR and is that TACARR allows for the innovation distribution to switch based on the regime. This may be an advantage over TARR by allowing TACARR to be more flexible and better capture underlying asymmetry. In addition, the upward and downward ranges potentially contain more information than range, thus allowing TACARR switch regimes more dynamically compared to TARR. Of course, this extra agility may not be needed when modeling a particular empirical series, resulting in a model that is too flexible. However, there can be instances when such flexibility is warranted. Thus, we propose TACARR as a potential alternative to TARR rather than a model that is superior to it.

The TACARR model can also be viewed as an asymmetric alternative to the ACARR and FACARR models. All three models treat price ranges asymmetrically, but TACARR has a lesser number of variables compared to the FACARR model. Therefore, the TACARR model is more parsimonious than FACARR. While TACARR and ACARR has the same number of parameters in the expression for the conditional range, the assumption that the upward and downward ranges propagate independently of one another without any cross-feedback between the two ranges may be unrealistic. Moreover, inspection of several price series show that upward and downward price range data contained excess number of zeros and thus some distributions such as lognormal distribution had to be discarded when using the ACARR model. However, in our proposed model, range data, which are almost always positive, is what is modeled, allowing one to consider distributions with positive support.

### 3.1. The ARMA Representation of The TACARR Model

This section presents the ARMA representation of the proposed TACARR model. Define the zero mean martingale difference process $\{\eta_t\}$ such that:

$$\eta_t = R_t - E(R_t \mid \mathbb{F}_{t-1}) = R_t - \lambda_t,$$
$$E(\eta_t) = E(R_t - \lambda_t) = 0, \text{ and}$$
$$\text{cov}(\eta_t, \eta_{t-h}) = E(\eta_t \eta_{t-h}) = 0 \text{ for } h \geq 1.$$

The proposed TACARR (*l*, *p*, *q*) model given in (3.1) can be rearranged as an ARMA process with order *k* and *q*, where $k = \max(p, q)$, as follows. We have,

$$R_t - \eta_t = \omega^{(m)} + \sum_{i=1}^{p} \alpha_i^{(m)} R_{t-i} + \sum_{j=1}^{q} \beta_j^{(m)} \left( R_{t-j} - \eta_{t-j} \right),$$

$$R_t = \omega^{(m)} + \sum_{i=1}^{p} \alpha_i^{(m)} R_{t-i} + \sum_{j=1}^{q} \beta_j^{(m)} R_{t-j} - \sum_{j=1}^{q} \beta_j^{(m)} \eta_{t-j} + \eta_t,$$

$$R_t = \omega^{(m)} + \sum_{i=1}^{k=\max(p,q)} \left( \alpha_i^{(m)} + \beta_i^{(m)} \right) R_{t-i} - \sum_{j=1}^{q} \beta_j^{(m)} \eta_{t-j} + \eta_t.$$

Note that $\alpha_i^{(m)} = 0$, for $i > p$ and $\beta_j^{(m)} = 0$, for $j > q$, where *m*= (*U*, *D*).

## 4. Parameter Estimation

In this section the Maximum Likelihood procedure is developed to estimate the proposed TACARR model parameters. Here, we considered two versions of the TACARR model based on the distribution of the innovation terms $\{\varepsilon_t\}$. In the Exponential TACARR (ETACARR) version, the innovations in the two-price range process are assumed to follow an exponential distribution with mean one, as expressed in equation (4.1). In the Lognormal TACARR (LNTACARR) version the innovations are assumed to follow a threshold lognormal distribution in which the parameters of the distribution are driven by market behavior as presented in equation (4.4). In later parts of this section, out-of-sample forecasting and performance evaluation methods are presented and discussed.

### 4.1. Parameter Estimation Method for Exponential TACARR (ETACARR) Model

Let $\left\{ \varepsilon_t^{(m)} \right\}$ be a sequence of independent and identically distributed exponential innovation term for a given market behavior *m*= (*U* =upward market, *D* =downward market) such that:

$$\varepsilon_t = \begin{cases} \varepsilon_t^{(U)} \sim i.i.d.\ \exp(1); E\left(\varepsilon_t^{(U)}\right) = 1 : C_{l,t}^{(U)} \geq C_{l,t}^{(D)} \\ \varepsilon_t^{(D)} \sim i.i.d.\ \exp(1); E\left(\varepsilon_t^{(D)}\right) = 1 : C_{l,t}^{(U)} < C_{l,t}^{(D)} \end{cases}.$$
(4.1)

The parameter vector $\Phi = \left(\omega^{(U)}, \alpha^{(U)}, \beta^{(U)}, \omega^{(D)}, \alpha^{(D)}, \beta^{(D)}\right)'$ can be estimated by using the conditional likelihood function and applying the maximum likelihood estimation (MLE) procedure.

### 4.1.1. The Log Likelihood Function for the ETACARR Model.

The conditional distribution of $R_t$ given $\left[\{R_k\}_{k=1}^{t-1}, \{R_k^u\}_{k=1}^{t-1}, \{R_k^d\}_{k=1}^{t-1}\right]$ can be expressed as follows:

$$f\left(R_t \mid \{R_k\}_{k=1}^{t-1}, \{R_k^u\}_{k=1}^{t-1}, \{R_k^d\}_{k=1}^{t-1}, \Phi\right) = \frac{1}{\lambda_t}\exp\left(-\frac{R_t}{\lambda_t}\right),$$

$$\lambda_t = \begin{cases} \lambda_t^{(U)} = \omega^{(U)} + \sum_{i=1}^{p}\alpha_i^{(U)} R_{t-i} + \sum_{j=1}^{q}\beta_j^{(U)}\lambda_{t-j} : C_{l,t}^{(U)} \geq C_{l,t}^{(D)} \\ \lambda_t^{(D)} = \omega^{(D)} + \sum_{i=1}^{p}\alpha_i^{(D)} R_{t-i} + \sum_{j=1}^{q}\beta_j^{(D)}\lambda_{t-j} : C_{l,t}^{(U)} < C_{l,t}^{(D)} \end{cases}.$$
(4.2)

Therefore, the conditional likelihood function $L\left(\Phi \mid \{R_t\}_{t=1}^{T}, \{R_t^u\}_{t=1}^{T-1}, \{R_t^d\}_{t=1}^{T-1}\right)$ and the conditional log likelihood function of the data $l\left(\Phi \mid \{R_t\}_{t=1}^{T}, \{R_t^u\}_{t=1}^{T-1}, \{R_t^d\}_{t=1}^{T-1}\right)$ can be derived as follows:

$$L\left(\Phi \mid \{R_t\}_{t=1}^{T}, \{R_t^u\}_{t=1}^{T-1}, \{R_t^d\}_{t=1}^{T-1}\right) = \prod_{t=\max(l,p,q)}^{T} f(R_t \mid \mathbb{F}_{t-1}, \Phi),$$

$$l\left(\Phi \mid \{R_t\}_{t=1}^{T}, \{R_t^u\}_{t=1}^{T-1}, \{R_t^d\}_{t=1}^{T-1}\right) = \ln\left[L\left(\Phi \mid \{R_t\}_{t=1}^{T}, \{R_t^u\}_{t=1}^{T-1}, \{R_t^d\}_{t=1}^{T-1}\right)\right]$$

$$= \sum_{t=\max(p,q,l)}^{T} \ln\left[f\left(R_t \mid \{R_t\}_{t=1}^{T}, \{R_t^u\}_{t=1}^{T-1}, \{R_t^d\}_{t=1}^{T-1}, \Phi\right)\right],$$

$$\Rightarrow l\left(\Phi \mid \{R_t\}_{t=1}^{T}, \{R_t^u\}_{t=1}^{T-1}, \{R_t^d\}_{t=1}^{T-1}\right) = -\sum_{t=\max(l,p,q)}^{T}\left(\ln(\lambda_t) + \frac{R_t}{\lambda_t}\right).$$
(4.3)

## 4.2. Parameter Estimation Method for Lognormal TACARR (LNTACARR) Model

In this section, we employed the lognormal distribution to model the error term in the model. Let $\{\varepsilon_t^{(m)}\}$ be the sequence of independent and identically distributed lognormal innovation term for a given market regimes $m=$ ($U$=upward market, $D$=downward market) such that:

$$\varepsilon_t = \begin{cases} \varepsilon_t^{(U)} \sim i.i.d.\, LN\left(-\dfrac{\theta_{(U)}^2}{2}; \theta_{(U)}^2\right); E\left(\varepsilon_t^{(U)}\right)=1 : C_{l,t}^{(U)} \geq C_{l,t}^{(D)} \\ \varepsilon_t^{(D)} \sim i.i.d.\, LN\left(-\dfrac{\theta_{(D)}^2}{2}; \theta_{(D)}^2\right); E\left(\varepsilon_t^{(D)}\right)=1 : C_{l,t}^{(U)} < C_{l,t}^{(D)} \end{cases},$$

(4.4)

The maximum likelihood estimation procedure is used to estimate the parameter vector $\Phi = \left(\omega^{(U)}, \alpha^{(U)}, \beta^{(U)}, \omega^{(D)}, \alpha^{(D)}, \beta^{(D)}, \theta_{(U)}, \theta_{(D)}\right)'$.

**4.2.1. The Log Likelihood Function for the LNTACARR Model.** The conditional distribution of $R_t$ given $\left[\{R_k\}_{k=1}^{t-1}, \{R_k^u\}_{k=1}^{t-1}, \{R_k^d\}_{k=1}^{t-1}\right]$ can be expressed as follows:

$$f\left(R_t \mid \{R_k\}_{k=1}^{t-1}, \{R_k^u\}_{k=1}^{t-1}, \{R_k^d\}_{k=1}^{t-1}, \Phi\right)$$

$$= \dfrac{1}{\sqrt{2\pi\theta_{(M_t)}^2}\, R_t} \exp\left[-\dfrac{1}{2\theta_{(M_t)}^2}\left(\ln(R_t) - \ln(\lambda_t) + \dfrac{\theta_{(M_t)}^2}{2}\right)^2\right],$$

$$\lambda_t = \begin{cases} \lambda_t^{(U)} = \omega^{(U)} + \sum_{i=1}^{p} \alpha_i^{(U)} R_{t-i} + \sum_{j=1}^{q} \beta_j^{(U)} \lambda_{t-j} : C_{l,t}^{(U)} \geq C_{l,t}^{(D)} \\ \lambda_t^{(D)} = \omega^{(D)} + \sum_{i=1}^{p} \alpha_i^{(D)} R_{t-i} + \sum_{j=1}^{q} \beta_j^{(D)} \lambda_{t-j} : C_{l,t}^{(U)} < C_{l,t}^{(D)} \end{cases}.$$

(4.5)

The conditional likelihood function $L\left(\Phi \mid \{R_t\}_{t=1}^{T}, \{R_t^u\}_{t=1}^{T-1}, \{R_t^d\}_{t=1}^{T-1}\right)$ and the conditional log likelihood function of the data $l\left(\Phi \mid \{R_t\}_{t=1}^{T}, \{R_t^u\}_{t=1}^{T-1}, \{R_t^d\}_{t=1}^{T-1}\right)$ can be derived as follows:

$$L\left(\Phi\mid\{R_t\}_{t=1}^{T},\{R_t^u\}_{t=1}^{T-1},\{R_t^d\}_{t=1}^{T-1}\right)=\prod_{t=\max(l,p,q)}^{T}f\left(R_t\mid\{R_k\}_{k=1}^{t-1},\{R_k^u\}_{k=1}^{t-1},\{R_k^d\}_{k=1}^{t-1},\Phi\right),$$

$$l\left(\Phi\mid\{R_t\}_{t=1}^{T},\{R_t^u\}_{t=1}^{T-1},\{R_t^d\}_{t=1}^{T-1}\right)=\ln\left[L\left(\Phi\mid\{R_t\}_{t=1}^{T},\{R_t^u\}_{t=1}^{T-1},\{R_t^d\}_{t=1}^{T-1}\right)\right]$$

$$=\sum_{t=\max(l,p,q)}^{T}\ln\left[f\left(R_t\mid\{R_k\}_{k=1}^{t-1},\{R_k^u\}_{k=1}^{t-1},\{R_k^d\}_{k=1}^{t-1},\Phi\right)\right],$$

$$l\left(\Phi\mid\{R_t\}_{t=1}^{T},\{R_t^u\}_{t=1}^{T-1},\{R_t^d\}_{t=1}^{T-1}\right)$$

$$=-\frac{1}{2}\sum_{t=\max(l,p,q)}^{T}\left[\ln(2\pi\theta_{(M_t)}^2)+2\ln(R_t)+\frac{1}{\theta_{(M_t)}^2}\left(\ln(R_t)-\ln(\lambda_t)+\frac{\theta_{(M_t)}^2}{2}\right)^2\right].$$

(4.6)

The parameters for the proposed model can be estimated by using the MLE method as discussed in the above section using in-sample data.

Next, we evaluate the in-sample performance of the proposed TACARR model with other conditional heteroscedastic range-based model by comparing the Root Mean Square Error (RMSE) and the Mean Absolute Error (MAE) values. Note that

$$RMSE=\sqrt{\frac{\sum_{t=1}^{N}\left(R_t-\hat{R}_t\right)^2}{N}};\ MAE=\frac{\sum_{t=1}^{N}\left|R_t-\hat{R}_t\right|}{N}.$$

Here, $R_t$ is the price range and $\hat{R}_t$ is the predicted price range at time period $t$, respectively. Moreover, $N$ is the in-sample size.

### 4.3. Out-Of-Sample Forecasting

We used a rolling window approach to forecast out-of-sample values, which was shown to produce better forecast by Zivot and Wang (2003). In this approach, we divided the entire sample period (sample size $=T$) into two periods namely in-sample period (in-sample size$=N<T$) and out-of-sample period. The first one-step-ahead out-of-sample forecasting is carried out using the all the $N$ in-sample data for model estimation. The method is described in detail bellow:

Let define, $R_{1:N}(1)$ be the one step ahead forecast of $R_{N+1}$ where $R_{N+1}=\lambda_{N+1}\varepsilon_{N+1}$. Then:

$$R_{1:N}(1) = E(R_{N+1} | \mathbb{F}_N) = \lambda_{N+1},$$

$$\lambda_{N+1} = \begin{cases} \lambda_{N+1}^{(U)} = \omega^{(U)} + \sum_{i=1}^{p} \alpha_i^{(U)} R_{N+1-i} + \sum_{j=1}^{q} \beta_j^{(U)} \lambda_{N+1-j} : C_{l,t}^{(U)} \geq C_{l,t}^{(D)} \\ \lambda_{N+1}^{(D)} = \omega^{(D)} + \sum_{i=1}^{p} \alpha_i^{(D)} R_{N+1-i} + \sum_{j=1}^{q} \beta_j^{(D)} \lambda_{N+1-j} : C_{l,t}^{(U)} < C_{l,t}^{(D)} \end{cases}.$$

(5.1)

Therefore, one step ahead forecast value $R_{1:N}(1)$ is calculated by using the conditional expectation of range given information up to time $N$.

After calculating the forecasted value for the $(N+1)^{th}$ observation the sample window is moved to $(2: N+1)$ to forecast $(N+2)^{th}$ observation. We considered the window of $(2: N+1)$ as the new in-sample data and recalculated the model parameters based on this new data. After the estimation, the estimated parameters were applied to the one step ahead forecasting method in equation (5.1) to calculate $R_{2:N+1}(1)$ which the forecasted value is for $R_{N+2}$. This process was repeated until all the future values in the out-of-sample data were forecast. The same rolling window procedure was employed for competitors to the proposed model. To check the forecasting accuracy of the proposed model with other competitive range models, the Diebold and Marino (DM) test was performed on the residuals (see Diebold & Marino, 1995).

## 5. Simulation Study

We investigated the finite sample performance of the MLE method using a simulation study. We used the R software to generate the relevant data. Length of the time series studied was set to $T = 1000$ and $T = 3000$, and $1000$ simulations runs were carried out for each sample size combination. We carried out the simulation study for two different distributions of error, namely the exponential and the lognormal. Simulations were carried out for both the exponential TACARR and the lognormal TACARR models. The simulation study was restricted to ETACARR (1, 1, 1) model and LNTACARR (1, 1, 1) models only, but we did consider several parameters combinations for these models. The simulation study consisted of two parts. First, we generated the price range data for the proposed ETACARR model and the LNTACARR model based on the equation (4.1) – (4.2) and (4.4) – (4.5), respectively. Then, we maximized the profile likelihood functions (4.3) and (4.6), for the ETACARR and LNTACARR models, respectively, using the constrained nonlinear optimization function *nloptr* in R. The Mean Absolute Deviation Error (MADE) was utilized as the evaluation criterion. The MADE is defined as, $\frac{1}{s}\sum_{j=1}^{s}|\hat{\phi}_j - \phi_j|$ where $s$ is the number of replications. Simulation results are reported in Table 1 and Table 2.

Table 1 presents the simulation study results for the ETACARR model, and results indicate that the MLE method did a good job in estimating model parameters. As expected, the accuracy increased with the size of the sample size. The simulation results presented in Table 2 also show that LNTACARR model parameters were estimated with higher accuracy using the MLE method. It can also be seen that MADE values decreased with the increase in sample size.

Table 1: Means of MLE estimates and MADE (within parentheses), for ETACARR model with order (1, 1, 1)

|  | **Upward Market Model** | | | **Downward Market Model** | | |
|---|---|---|---|---|---|---|
|  | $\omega^{(U)}$ | $\alpha^{(U)}$ | $\beta^{(U)}$ | $\omega^{(D)}$ | $\alpha^{(D)}$ | $\beta^{(D)}$ |
| True Value | 0.01 | 0.10 | 0.80 | 0.10 | 0.20 | 0.70 |
| $T$=1000 | 0.0170 (0.0152) | 0.0977 (0.0253) | 0.7864 (0.0449) | 0.0996 (0.0248) | 0.1982 (0.0340) | 0.7045 (0.0599) |
| $T$=3000 | 0.0130 (0.0101) | 0.0995 (0.0142) | 0.7943 (0.0283) | 0.1003 (0.0153) | 0.2004 (0.0196) | 0.6995 (0.0367) |
| True Value | 0.01 | 0.30 | 0.60 | 0.10 | 0.20 | 0.50 |
| $T$=1000 | 0.0132 (0.0096) | 0.3011 (0.0379) | 0.5867 (0.0583) | 0.1001 (0.0170) | 0.1964 (0.0407) | 0.5033 (0.0769) |
| $T$=3000 | 0.0110 (0.0058) | 0.3007 (0.0223) | 0.5955 (0.0349) | 0.1008 (0.0098) | 0.2006 (0.0239) | 0.4965 (0.0443) |
| True Value | 0.05 | 0.15 | 0.50 | 0.10 | 0.20 | 0.30 |
| $T$=1000 | 0.0551 (0.0247) | 0.1542 (0.0395) | 0.4702 (0.1581) | 0.0985 (0.0261) | 0.1959 (0.0462) | 0.3122 (0.1582) |
| $T$=3000 | 0.0517 (0.0126) | 0.1505 (0.0221) | 0.4897 (0.0809) | 0.1008 (0.0154) | 0.2015 (0.0264) | 0.2944 (0.0933) |

Table 2: Means of MLE estimates and MADE (within parentheses), for LNTACARR model with order (1, 1, 1)

| T=3000 | T=1000 | True Parameter | T=3000 | T=1000 | True Parameter | T=1000 | T=3000 | True Parameter | | |
|---|---|---|---|---|---|---|---|---|---|---|
| 0.0490 (0.0077) | 0.0502 (0.0122) | 0.05 | 0.0106 (0.0054) | 0.0125 (0.0085) | 0.01 | 0.0131 (0.0109) | 0.0110 (0.0075) | 0.01 | $\omega^{(U)}$ | Upward Market |
| 0.1525 (0.0209) | 0.2080 (0.0597) | 0.15 | 0.3002 (0.0235) | 0.2974 (0.0394) | 0.30 | 0.0983 (0.0178) | 0.0998 (0.0102) | 0.10 | $\alpha^{(U)}$ | |
| 0.5037 (0.0514) | 0.4430 (0.0928) | 0.50 | 0.5974 (0.0338) | 0.5906 (0.0551) | 0.60 | 0.7950 (0.0318) | 0.7980 (0.0204) | 0.80 | $\beta^{(U)}$ | |
| 0.0898 (0.0026) | 0.0911 (0.0046) | 0.09 | 0.9968 (0.0279) | 0.9924 (0.0489) | 1.00 | 0.2483 (0.0125) | 0.2491 (0.0071) | 0.25 | $\theta^2_{(U)}$ | |
| 0.1008 (0.0071) | 0.0997 (0.0111) | 0.10 | 0.1002 (0.0092) | 0.1005 (0.0157) | 0.10 | 0.1000 (0.0195) | 0.1002 (0.0126) | 0.10 | $\omega^{(D)}$ | Downward Market |
| 0.2136 (0.0200) | 0.2009 (0.0298) | 0.20 | 0.1995 (0.0233) | 0.1991 (0.0410) | 0.20 | 0.1997 (0.0289) | 0.2000 (0.0166) | 0.20 | $\alpha^{(D)}$ | |
| 0.2824 (0.0468) | 0.3007 (0.0727) | 0.30 | 0.5000 (0.0428) | 0.5007 (0.0756) | 0.50 | 0.7016 (0.0482) | 0.7002 (0.0302) | 0.70 | $\beta^{(D)}$ | |
| 0.0401 (0.0011) | 0.0398 (0.0021) | 0.04 | 0.9993 (0.0274) | 0.9947 (0.0516) | 1.00 | 0.6379 (0.0317) | 0.6401 (0.0171) | 0.64 | $\theta^2_{(D)}$ | |

# 6. Empirical Results

## 6.1. The Data Set

IBM stock indices were used to gauge the performance of the proposed TACARR model and to compare it with competitor models. The sample periods for the IBM data spanned from January 01, 2002, to March 13, 2020. Daily values for the opening price, closing price, high price, low price, and adjusted price were reported over the span of the study period. The data set was obtained from the Yahoo Finance (https://finance.yahoo.com/) using the *quantmod* package in R. The data set was divided in to two sub samples: the first sub sample was considered the in-sample period and this sample was used to estimate the model parameters and in-sample predictions. In-sample periods for IBM spanning from January 01, 2002, to December 31, 2019. The second sub sample, called as the out-of-sample period was used for out-of-sample forecasting. Out-of-sample periods for IBM elapsed from January 1, 2020, to March 13, 2020. Table 3 presents the summary statistics of the IBM price range data, which was calculated as given in (2.1).

Summary statistics for the IBM stock index is presented in the Table 3. As seen in the table, the high persistence of volatility is present for this stock. For an example, the Ljung-Box test statistic results for lags 1, 5 and 22 show that all three forms of ranges exhibit highly significant serial correlations. The downward component has higher Ljung-Box test statistic values for all lags than the upward price range component, indicating that higher persistence exists in the downward price range series. The price range data did not contain zero values for the range, but upward range and downward range components had a considerable number of zero-valued observations. Positive skewness and absence of zeros in range data suggested that a probability density function (pdf) with positive support, such as a lognormal can be used. By contrast, upward range, and downward range components, while positively skewed, have notable number of observations of zeros. This implies a positively skewed but a pdf with support that includes zero, such as an exponential distribution, or even one that includes positive probability weight at zero, must be used when modeling these components with the ACARR and FACARR models. However, such restrictions do not apply for the TACARR model.

Table 3: Summary statistics of the IBM price range data

| Statistics | Price Range | Upward Price Range | Downward Price Range |
|---|---|---|---|
| Number of Days | 4581 | 4581 | 4581 |
| Minimum | 0.2928 | 0.0000 | 0.0000 |
| Mean | 1.6834 | 0.8717 | 0.8118 |
| Maximum | 11.2642 | 8.0510 | 8.4991 |
| Standard Deviation | 1.0774 | 0.8279 | 0.8741 |
| Skewness | 2.7976 | 2.4293 | 2.7510 |
| Number of Zeros | 0 | 107 | 160 |
| Q (1)* | 1891.7*** | 235.45*** | 467.34*** |
| Q (5)* | 8013*** | 923.36*** | 2954*** |
| Q (22)* | 25498*** | 3271.1*** | 5915*** |

Note: *Q($h$) indicates the Ljung-Box statistic for lag $h$. *** indicates 1% significance level.

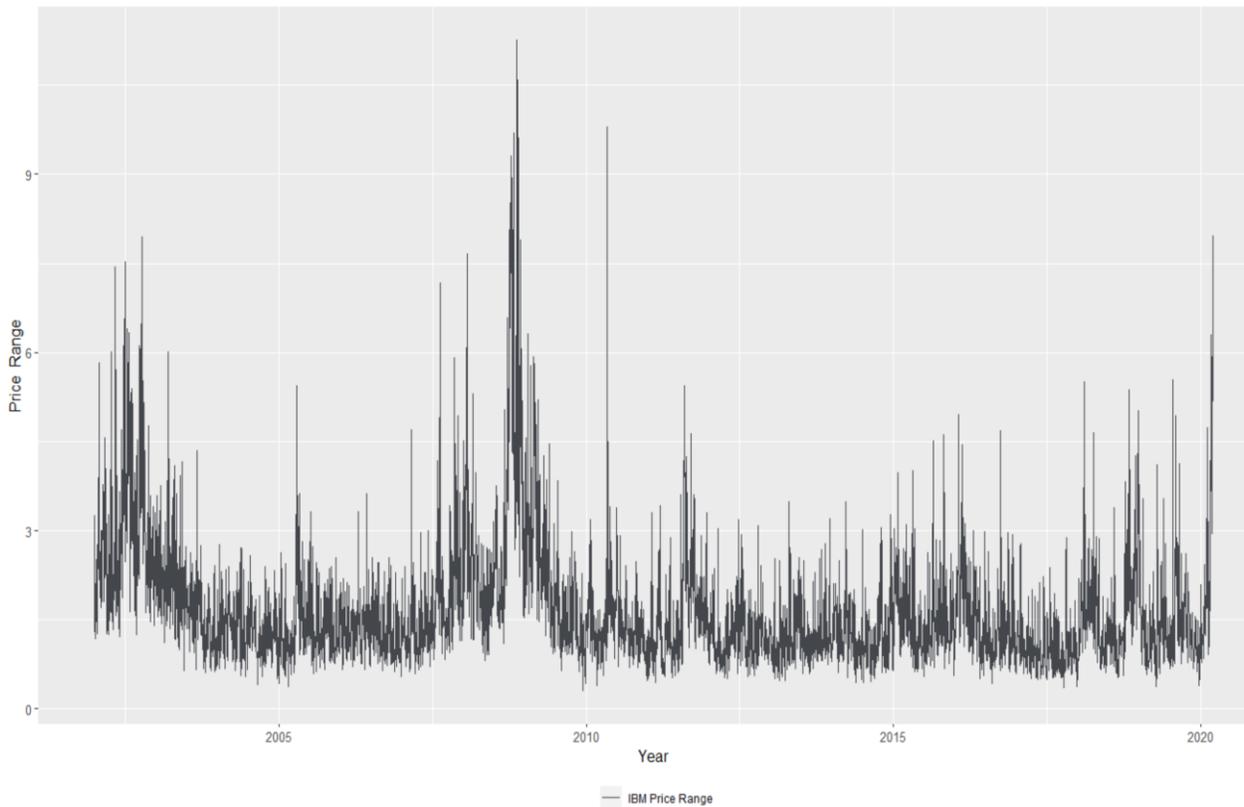

Figure 1: IBM price range data for the period of 01/01/2002 to 03/13/2020

As shown in Figure 1, for the IBM price range data, the price volatility was high at the beginning of the sample period and then decreased. Moreover, daily price volatility fluctuated rapidly during 2007-2009 recession. After the recession ended volatility dropped but it seems to have picked up after the first quarter of 2020, most probably due to the Covid-19 pandemic and its influence on the financial market.

## 6.2. In-Sample Estimation Results

We introduced two versions of the TACARR models, differentiated by the distribution of the innovations. The Kolmogorov-Simonov (KS) test was employed to identify which of the two models better fit the price range data. The KS compared whether the standardized residual series followed the reference distribution. In addition, we calculated the Log Likelihood Function (LLF), Akaike Information Criteria (AIC), Bayesian Information Criteria (BIC) for each model. The model with smaller AIC, BIC values and larger LLF value is considered to be a better model than the other. Furthermore, diagnostic tests, such as the Ljung-Box test were conducted to check whether residuals were independent and identically distributed.

Parameter estimation results for the IBM price range data for the ETACARR and LNTACARR are summarized in the Table 4, and it contains two panels where upper panel (A) presents the parameter estimation results and model selection statistics such as LLF, AIC and BIC values. The lower panel (B) summarizes the diagnostic test results for the standardized residuals. The KS tests for the exponential and lognormal cases were separately considered to identify whether the residual series followed the hypothesized distribution with the estimated parameters. Then Ljung-Box test statistics for 1, 5 and 22, lags were used to check whether the residual series are exhibited any serial correlations. These are reported in Table 4 in rows denoted by Q (1), Q(5), and Q (22) with the respective *p*-values.

Results in Table 4 show that the Lognormal TACARR (*l*,1,1) models, for all three values of *l*, have smaller AIC, BIC values and larger LLF values when compared to its exponential alternative. Thus, it suggests that the LNTACRR models fit the data better than its exponential alternative. The LNTACARR models indicate that upward and downward markets have different variance parameters ($\theta^2_{(U)} \neq \theta^2_{(D)}$). The persistence estimate for the downward market is higher than that of the upward market ($\alpha^{(U)} + \beta^{(U)} < \alpha^{(D)} + \beta^{(D)}$). This shows that the downward market was more volatile and persistent than the upward market. Moreover, KS tests for the LNTACRR models suggest that the standardized residuals followed the hypothesized threshold lognormal distribution. However, for the ETACARR models, the standardized residuals did not follow the proposed threshold exponential distribution according to the KS test results. Based on the *p*-values for the Ljung-Box test for lags 1, 5, and 22, it can be concluded that we failed to reject the null hypothesis at the 0.01 significance level, suggesting that residuals are not serially correlated. Therefore, it can be concluded that the LNTACARR model fits better than the ETACARR model for all three of the *l* values. Dues to the clearly superior performance of the LNTACARR model, we will concentrate on this model from here on. It should be noted, however, that this better performance is detected in relation to the IBM stock data and that a similar comparison to the above should be carried out for any new data set.

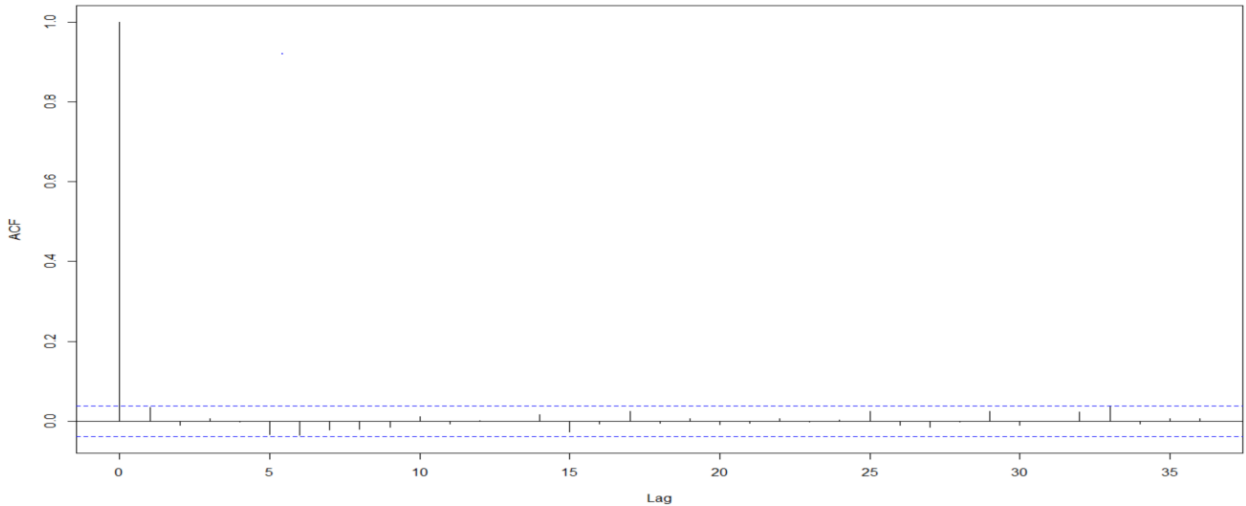

Figure 2: ACF plot of the LNTACARR (1, 1, 1) residuals

The ACF plot indicates that all residuals of the LNTACARR (1, 1, 1) were within the 99% confidence interval. Therefore, graphically, it is seen that residuals are independent and identically distributed.

Table 4: Estimation and diagnostic test results of the TACARR models with exponential and lognormal disturbance term for IBM data

| Table 4A: Estimation of the TACARR model (standard errors) | | | | | | |
|---|---|---|---|---|---|---|
| | TACARR (22,1,1) | | TACARR (5,1,1) | | TACARR (1,1,1) | |
| | Exponential | Lognormal | Exponential | Lognormal | Exponential | Lognormal |
| $\omega^{(U)}$ | 0.0569 | 0.0535 | 0.0749 | 0.0684 | 0.0855 | 0.0798 |
| | (0.0324) | (0.0119) | (0.0359) | (0.0129) | (0.0493) | (0.0181) |
| $\alpha^{(U)}$ | 0.1680 | 0.1683 | 0.1776 | 0.1674 | 0.1615 | 0.1573 |
| | (0.0432) | (0.0164) | (0.0428) | (0.0155) | (0.0395) | (0.0147) |
| $\beta^{(U)}$ | 0.7944 | 0.7960 | 0.7682 | 0.7831 | 0.7672 | 0.7777 |
| | (0.0544) | (0.0207) | (0.0543) | (0.0197) | (0.0553) | (0.0206) |
| $\theta^2_{(U)}$ | | 0.1316 | | 0.1361 | | 0.1386 |
| | | (0.0039) | | (0.0039) | | (0.0040) |
| $\omega^{(D)}$ | 0.0554 | 0.0488 | 0.0364 | 0.0321 | 0.0313 | 0.0248 |
| | (0.0316) | (0.0114) | (0.0366) | (0.0135) | (0.0547) | (0.0198) |
| $\alpha^{(D)}$ | 0.2301 | 0.2175 | 0.2169 | 0.2138 | 0.2373 | 0.2291 |
| | (0.0455) | (0.0168) | (0.0438) | (0.0163) | (0.0456) | (0.0166) |
| $\beta^{(D)}$ | 0.7374 | 0.7535 | 0.7680 | 0.7720 | 0.7627 | 0.7709 |
| | (0.0562) | (0.0207) | (0.0559) | (0.0209) | (0.0631) | (0.0230) |
| $\theta^2_{(D)}$ | | 0.1411 | | 0.1359 | | 0.1313 |
| | | (0.0042) | | (0.0040) | | (0.0040) |
| LLF | -6547.26 | -3609.47 | -6546.68 | -3605.79 | -6544.64 | -3592.76 |
| AIC | 13106.52 | 7234.95 | 13105.36 | 7227.57 | 13101.27 | 7201.51 |
| BIC | 13145.03 | 7286.30 | 13143.87 | 7278.92 | 13139.79 | 7252.86 |
| Table 4B: Diagnostic test results of the TACARR model (p-values) | | | | | | |
| Residual Tests | TACARR (22,1,1) | | TACARR (5,1,1) | | TACARR (1,1,1) | |
| | Exponential | Lognormal | Exponential | Lognormal | Exponential | Lognormal |
| KS | 0.3578 | 0.0254 | 0.3593 | 0.0274 | 0.3609 | 0.0252 |
| | (0.0000) | (0.1080) | (0.0000) | (0.0672) | (0.0000) | (0.1136) |
| Q(1) * | 4.615 | 6.2995 | 5.2874 | 7.014 | 3.7791 | 5.2215 |
| | (0.3169) | (0.0121) | (0.0215) | (0.0081) | (0.0519) | (0.0223) |
| Q(5) * | 10.672 | 12.26 | 11.805 | 13.271 | 9.8418 | 11.281 |
| | (0.0583) | (0.0315) | (0.0376) | (0.0210) | (0.0799) | (0.0461) |
| Q(22)* | 30.55 | 33.61 | 32.457 | 35.575 | 28.95 | 31.66 |
| | (0.1102) | (0.0538) | (0.0699) | (0.0337) | (0.1463) | (0.0834) |

Note: *Q(*h*) indicates the Ljung-Box statistic for lag *h*

Results in Table 4 can also be used to determine the optimal lag ($l$) one should select from among the choices 1, 5, and 22, for the proposed LNTACARR model. The optimal lag selection is an important task because it decides how many previous periods (i.e., day, week, or months) that we need to consider for categorizing the market. Here we considered three different lag values: 1, 5 and 22. In the case of $l=1$ the status of the market regime was determined based on the previous days upward and downward range data. Put simply, $l=1$, considered the most recent financial information to decide the market regime. When $l=5$, market is segregated based on the volatility information of the past business week, while the case of $l=22$ divided the market into two regimes based on the market information from the last business month. (Please note that 5 days is considered as one business week, and 22 days is equal to one business month). In this study, we selected the LNTACARR ($l$, 1, 1) model with lowest AIC, BIC and largest LLF values. Based on results in Table 4, the LNTACARR (1, 1, 1) model has the lowest AIC and BIC values and also largest LLF value; therefore, for the IBM data $l=1$, is the optimal lag value. This can be interpreted as implying that the LNTACARR model with $l=1$ is more sensitive to the market information through a thresholding mechanism based on previous days market behavior. It is also important to note that the proposed model addresses the asymmetric behavior in the financial market through this thresholding process.

To summarize, among all the TACARR ($l$, 1, 1) models, we investigated in the previous section, we selected LNTACARR (1, 1, 1) model as our candidate for modeling the IBM stock data. The next step is to compare and contrast the adequacy of this model with other conditional heteroscedastic range-based models. The fitted LNTACARR model we obtained is as follows:

$$\lambda_t = \begin{cases} \lambda_t^{(U)} = 0.0798 + 0.1573 R_{t-1} + 0.7777 \lambda_{t-1} : R_{t-1}^u \geq R_{t-1}^d \\ \lambda_t^{(D)} = 0.0248 + 0.2291 R_{t-1} + 0.7709 \lambda_{t-1} : R_{t-1}^d < R_{t-1}^u \end{cases},$$

$$\varepsilon_t = \begin{cases} \varepsilon_t^{(U)} \sim LN(-0.0693; 0.1386) : R_{t-1}^u \geq R_{t-1}^d \\ \varepsilon_t^{(D)} \sim LN(-0.0656; 0.1313) : R_{t-1}^u < R_{t-1}^d \end{cases}.$$

To gauge the in-sample performance of the chosen LNTACARR (1, 1, 1) model we considered Root Mean Squared Error (RMSE) and Mean Absolute error (MAE) values. We used these statistics to compare it with alternative range models, such as the LNCARR (1, 1), ACARR (1, 1), FACARR (1, 1) and LNARR (1, 1). Since our study showed that upward and downward ranges had large numbers of zeroes for the IBM price range data, the lognormal distribution cannot be considered for modeling upward and downward price range components in the competing ACARR and FACARR models. Therefore, it was assumed that upward and downward innovation terms are distributed exponentially in the ACARR and the FACARR models. In addition to comparing the performance of the models over the complete span of the data, we also compared the using only the data for the 2007-2009 recession period.

Based on the comparison results presented in the Table 5, the proposed LNTACARR model had the lowest RMSE and MAE values for the in-sample period when compared to that of the other candidate models. This result suggests that with respect to the model fit, the proposed model performed slightly better than the others. Moreover, during the economic recession period, the lowest RMSE value was recorded for LNTACARR model. However, MAE value was lower in the FACARR model. Overall, it appears that the lognormal version of the proposed TACARR formulation is, in general, better suitable to analyze the volatility data compared to the LNCARR (1, 1), ACARR (1, 1), FACARR (1, 1) and LNTARR (1, 1) models.

Table 5: In-sample comparison between LNCARR (1, 1), ACARR (1, 1), FACARR (1, 1), LNTARR (1, 1) and LNTACARR (1, 1) for IBM data

| Table 5A: Model performance comparison during full in-sample period | | | | | |
|---|---|---|---|---|---|
| Statistic | LNCARR (1,1) | ACARR (1,1) | FACARR (1,1) | LNTARR (1,1) | LNTACARR (1,1,1) |
| RMSE | 0.7289 | 0.7568 | 0.7241 | 0.7276 | **0.7224** |
| MAE | 0.4989 | 0.5139 | 0.4969 | 0.4966 | **0.4946** |
| Table 5B: Model performance comparison during economic recession period | | | | | |
| Statistic | LNCARR (1,1) | ACARR (1,1) | FACARR (1,1) | LNTARR (1,1) | LNTACARR (1,1,1) |
| RMSE | 0.9347 | 0.9396 | 0.9233 | 0.9371 | **0.9224** |
| MAE | 0.6970 | **0.6815** | 0.6890 | 0.6985 | 0.6890 |

## 6.3. Out-Of-Sample Forecasting

Out-of-sample performance of the proposed Lognormal TACARR (1, 1, 1) model was compared with four other models namely the LNCARR (1, 1), ACARR (1, 1), FACARR (1, 1), and the LNTARR (1, 1). In this study, the out-of-sample period started on January 01, 2020 and ended on March 13, 2020. Length of the out-of-sample period equals to 50 days. The out-of-sample period showed high volatility possibly due to the impact of the Covid-19 Pandemic on the financial market. First, the RMSE and MAE values were calculated and used these values as the performance indicator to gauge the performance of the proposed model. These results are presented in the Table 6. Then, we considered the Diebold & Marino (DM) test to check whether the proposed LNTACARR model with order (1, 1, 1) had a better forecasting accuracy than the other competitive models. In this test, the null hypothesis was that the LNTACARR (1, 1, 1) model had lower forecasting accuracy. The alternative hypothesis was stated that the one step ahead forecasted value of the LNTACARR (1, 1, 1) model was more accurate than the forecast values of its competitive model. The DM test result is summarized in Table 7.

Table 6: Out-of-sample comparison between LNCARR (1, 1), ACARR (1, 1), LNTARR (1, 1) and LNTACARR (1, 1, 1) for IBM data

| Statistic | LNCARR (1,1) | ACARR (1,1) | FACARR (1,1) | LNTARR (1,1) | LNTACARR (1,1,1) |
|---|---|---|---|---|---|
| RMSE | 1.2720 | 1.5203 | 1.2205 | 1.2820 | **1.1858** |
| MAE | 0.8371 | 0.9414 | 0.8024 | 0.8437 | **0.7752** |

According to Table 6, the LNTACARR model with order (1, 1, 1) had the lowest RMSE and MAE when compared to the other four models. Therefore, based on these accuracy measurements, it was concluded that proposed model performs better than LNCARR, ACARR, FACARR and LNARR models.

According to Table 7, it was concluded with 95% confidence that the propose LNTACARR model had higher forecasting accuracy than the other asymmetric range-based heteroscedastic models, such as LNCARR, ACARR and LNTARR.

Table 7: Diebold & Marino (DM) test results on IBM out-of-sample data

| Null Hypothesis | DM test statistics (p value) |
|---|---|
| Forecast LNCARR (1,1) model is same or more accurate than that of the LNTACARR (1,1) | -2.4721 (0.0067) |
| Forecast ACARR (1,1) model is same or more accurate than that of the LNTACARR (1,1) | -2.7660 (0.0028) |
| Forecast FACARR (1,1) model is same or more accurate than that of the LNTACARR (1,1) | -2.2861 (0.0112) |
| Forecast LNTARR (1,1) model is same or more accurate than that of the LNTACARR (1,1,1) | -2.7505 (0.0032) |

## 7. Conclusions

A Threshold Asymmetric Conditional Autoregressive Range (TACARR) model was proposed to account for asymmetric impact of market forces on volatility. It is a range-based formulation for modeling and forecasting financial price range data proposed as an alternative to other range-based models. The asymmetric impact on volatility is achieved through a thresholding mechanism that switches the model between two regimes, namely upward market, and downward market, based on the number of times upmarket swings dominated down market swings over a prescribed past period. This switching occurs dynamically, without using a pre-determined threshold, and is based on observed components of the price range data. In addition, the disturbance term in the model to behave differently in each market regime. It is proposed as an alternative approach for capturing the asymmetric behavior in the market by adjusting the threshold value according to the market conditions related to the asset under study. The proposed model has two versions, namely ETACARR and LNTACARR, differentiated based on the distribution of the underlying innovations. An empirical study using IBM price range data shows that the LNTACARR model performed better than its exponential alternative across all the different periods considered for making switching decisions. Moreover, we found that switching based on previous day's upward and downward ranges provided the best fit. For the in-sample and recession data, the predicted values of LNTACARR (1, 1, 1) model had better fit when compared to the LNCARR, ACARR, FACARR and LNTARR models. Diagnostic test results for the selected model suggested that the residuals were independent and identically distributed, and it followed a lognormal distribution. Finally, out-of-sample forecasting evaluation was considered and according to the Root Mean Squared Error and Mean Absolute error, the LNTACARR performed slightly better than the other four models. Furthermore, the Diebold-Merino test for the forecasting accuracy indicated that the proposed model had more accurate forecast than LNCARR, ACARR, FACARR and LNTARR models for the IBM price range data.

Tong, H., (1978). On a threshold model. In Pattern recognition and Signal Processing (ed. C.H. Chen), pp. 575-586. Amsterdam: Sijthoff and Noordhoff.

Tsay, R. (1989). Testing and Modeling Threshold Autoregressive Processes. Journal of the American Statistical Association, 84(405), 231-240.

Xie H. (2018). Financial volatility modeling: The feedback asymmetric conditional autoregressive range model. Journal of Forecasting. 2019; 38:11–28.

Xie, H. B., & Wu, X. Y. (2017). A conditional autoregressive range model with gamma distribution for financial volatility modeling. Economic Modelling, 64, 349–356.

Zakoian, J. M. (1991). Threshold heteroskedastic models, Technical report, INSEE.

Zakoïan, J.M. (1994). Threshold Heteroskedastic Models. Journal of Economic Dynamics and Control, 18, 931-955.

Zhang, M. Y., Russell, J., & Tsay, R. S. (2001). A nonlinear autoregressive conditional duration model with applications to financial transaction data. Journal of Econometrics, 104,179–207.

Zivot, E., & Wang, J. (2003). Rolling analysis of time series. In *Modeling Financial Time Series with S-Plus®* (pp. 299-346). Springer, New York, NY.
33